\documentclass[a4paper]{article}

\usepackage[english]{babel}
\usepackage[utf8]{inputenc}
\usepackage[T1]{fontenc}

\usepackage{siunitx}
\usepackage[a4paper,top=3cm,bottom=2cm,left=3cm,right=3cm,marginparwidth=1.75cm]{geometry}

\usepackage{amssymb}
\usepackage{amsmath}
\usepackage{graphicx}
\usepackage{url}
\usepackage[colorlinks=true, allcolors=blue]{hyperref}
\usepackage{subfigure}
\usepackage{braket}
\usepackage{authblk}

\newcommand{\ignore}[1]{}
\DeclareUnicodeCharacter{2212}{-}

\numberwithin{equation}{subsection}

\title{\textbf{Magnetic-Field-Dependent Stimulated Emission from Nitrogen-Vacancy Centres in Diamond}}
\author[1,\footnote{felix.hahl@iaf.fraunhofer.de}]{F. Hahl}
\author[1]{L. Lindner}
\author[1]{X. Vidal}
\author[2]{T. Ohshima}
\author[2]{S. Onoda}
\author[2]{S. Ishii}
\author[3,4]{A. M. Zaitsev}
\author[5]{M. Capelli}
\author[1]{T. Luo}
\author[6]{B. C. Gibson}
\author[6]{A. D. Greentree}
\author[1, \footnote{jan.jeske@iaf.fraunhofer.de}]{J. Jeske}

\affil[1]{Fraunhofer Institut für Angewandte Festkörperphysik (IAF), Tullastrasse 72, 79108 Freiburg, Germany}
\affil[2]{National Institutes for Quantum and Radiological Science and Technology (QST), 1233
Watanuki, Takasaki, Gunma 370-1292, Japan}
\affil[3]{The College of Staten Island/CUNY, 2800 Victory Blvd., Staten Island, NY 10312, USA}
\affil[4]{Gemological Institute of America, 50 W 47th St \#800, New York, NY 10036, USA}
\affil[5]{School of Science, RMIT University, Melbourne, VIC 3001, Australia}
\affil[6]{ARC Centre of Excellence for Nanoscale BioPhotonics, School of Science, RMIT University, Melbourne, VIC 3001, Australia}

\begin{document}
\graphicspath{{pictures/}}
\selectlanguage{english}
\maketitle

\begin{abstract}
Negatively charged nitrogen-vacancy centres in diamond are promising quantum magnetic field sensors. Laser threshold magnetometry has been a theoretical approach for the improvement of NV-centre ensemble sensitivity via increased signal strength and magnetic field contrast. In this work we experimentally demonstrate laser threshold magnetometry. We use a macroscopic high-finesse laser cavity containing a highly NV-doped and low absorbing diamond gain medium that is pumped at $\SI{532}{nm}$ and resonantly seeded at $\SI{710}{nm}$. This enables amplification of the signal power by stimulated emission of $\SI{64}{\%}$. We show the magnetic-field dependency of the amplification and thus, demonstrate magnetic-field dependent stimulated emission from an NV-centre ensemble. This emission shows a record contrast of $\SI{33}{\%}$ and a maximum output power in the mW regime. These advantages of coherent read-out of NV-centres pave the way for novel cavity and laser applications of quantum defects as well as diamond NV magnetic field sensors with significantly improved sensitivity for the health, research and mining sectors.
\end{abstract}

%###########################################################################################################
\section{Introduction}
Quantum magnetic field sensors are of increasing importance across a diverse range of fields. Particular examples include the monitoring and diagnosis of medical conditions through the examination of neuronal activity (magnetoencephalography , MEG) or cardiological signals (magnetocardiography, MCG)  \cite{Boto2018, Zhang2020a, Bison2009,Lembke2014}, in mining industry for geological exploration of  magnetic minerals and magnetic anomaly detection (MAD) \cite{Hall2014} or in fundamental studies of magnetism \cite{McCullian2020}.\par

Currently the best magnetic field sensitivities are reached by superconducting quantum interference devices (SQUIDS) and vapour cells, like spin exchange relaxation-free (SERF) magnetometers reaching sensitivities below $\SI{1}{fT/\sqrt{Hz}}$ \cite{Kominis2003, Dang2010, Shah2013}. However SQUIDs require cryogenic cooling with liquid helium and vapour cells can only be operated in zero-field environments and require heating.\par 

The use of negatively charged nitrogen-vacancy (NV) centres for quantum magnetic sensing has caused great attention since it forms a quantum system that can be operated under ambient conditions in the earth's magnetic field or other background fields avoiding the need for cryogenic cooling and shielding, such as required for SQUIDS and vapour cells \cite{Kominis2003,Dang2010, Griffith2010}. Present sensitivities of NV-centre ensembles reach $\SI{}{pT/\sqrt{Hz}}$ \cite{Wolf2015, Zhang2020}, restricted by fluorescence collection efficiency, low contrast and reduced spin coherence time in highly doped diamond \cite{Degen2017, Taylor2008}. The improvement of the NV ensemble sensitivity to magnetic fields would enable versatile and robust room temperature sensing applications.\par

The unique optical and spin properties are due to the isolated NV-centre's electronic energy levels in the band-gap of diamond, which enable optical excitation, spin initialisation and detection. This allows for sensing temperature, strain and pressure, electric and magnetic fields with single or ensembles of NV-centres \cite{Doherty2014a, Neumann2013, Michl2019}. The photoluminescence (PL) differs between the "brighter" spin $m_{s}=0$ and "darker", $m_{s}=\pm1$ state. However, the maximum contrast achieved by single NV-centres up to date is $\SI{30}{\%}$ \cite{Jelezko2006} and for ensembles of NV-centres typically only $\thicksim\SI{5}{\%}$  because of the stray signal and inhomogeneous broadening from other defects leading to a decreasing signal-to-noise ratio \cite{Rondin2014} which can only slightly be improved to $\SI{9}{\%}$ by photocurrent detection of magnetic resonance (PDMR) \cite{Bourgeois2017}. Current NV-magnetometry is realised by collecting the PL of the NV-centre, both for single as well as for ensembles of NV-centres. To enhance the detection signal, the spontaneous emission that is sent out into the whole spatial angle is partially collected via optical lenses \cite{Rondin2014,Zhang2021, Kaupp2016}. In addition to limited PL collection efficiency the theoretical maximum of the contrast of an ensemble of NV-centres considering all four possible NV-directions \cite{Doherty2013} is $\SI{22}{\%}$ (see Supplementary).\par

A boost for the sensitivity is promised by laser threshold magnetometry (LTM) \cite{Jeske2016}, due to the competition between spontaneous and stimulated emission. This leads to an expected contrast of almost unity at the lasing threshold. Additionally, the coherently emitted photons by stimulated emission in an NV-laser are directional and lead to much higher signal collection efficiency compared to spontaneous emission. The predicted sensitivity of $\SI{1}{fT/\sqrt{Hz}}$  \cite{Jeske2016} for LTM would enable to overcome the three orders of magnitude between the predicted and state-of-the-art NV-ensemble sensitivities. Further theoretical concepts for laser threshold magnetometry using the NV's IR transition \cite{Dumeige2011}, a combination with diamond Raman lasers \cite{Nair2021} or visible absorption \cite{Webb2021} have been explored. Stimulated emission from NV-centres has been shown \cite{Jeske2017} as well as amplification by stimulated emission in a fibre cavity \cite{Nair2020}. For this reason a laser out of NV-centres is the goal to enable crossing boundaries in NV-centres sensitivities. However, despite several studies \cite{Mironov2021,Fraczek2017,Savvin2021,Nair2020,Savitski2017,Hacquebard2018,Dumeige2019,Nair2021,Webb2021} very strong stimulated emission signals or magnetic-field dependent coherent read-out have not been achieved so far.\par

Here we show magnetic-field dependent light amplification by stimulated emission in an ensemble of NV centres in a high-finesse cavity.  The NV centres were pumped at 532 nm, and the cavity seeded at 710 nm.  The cavity output at 710 nm increased by 64\% when pumped by the green laser.  By applying a permanent magnet to the NV ensemble, we observed a contrast (magnetic field to no-magnetic field) of 33\%.  This exceeds the theoretical maximum achievable for conventional ensemble photoluminescence. Furthermore, we show a new effect that we call induced absorption as an additional loss channel in the diamond medium which reduces the amplification and the cavity finesse at pump intensities $\gtrsim\SI{10}{kW/cm^{2}}$. The induced absorption prevents the NV-centres from self-sustained lasing activity. We show optically detected magnetic resonance (ODMR) with a contrast of 17\% which is higher than the simultaneously detected contrast of the PL reaching 11\%. We reach a shot-noise limited DC sensitivity of $\SI{14.6\pm1.3}{pT/\sqrt{Hz}}$ with a coherent laser signal output in the mW range. This demonstrates experimentally the principle of laser threshold magnetometry (LTM) for the first time.

%############################################################################################################################################################
\section{Results}
\subsection{High finesse NV-lasing cavity}
The experimental setup in Fig. 1a is designed to study the amplification by stimulated emission and the finesse of the cavity, simultaneously. To achieve strong stimulated emission, pumping of a high number of NV-centres is needed. For this, we use a highly NV-doped ($\approx\SI{1.8}{ppm}$) bulk diamond plate of thickness $l=\SI{295}{\mu m}$. We pump the NV-centres at $\lambda_{pump}=\SI{532}{nm}$ with a diode pumped solid state laser and a maximum output power of $\SI{12}{W}$. A macroscopic pumping volume is achieved by choosing the fundamental TEM$_{00}$ cavity mode to have a beam waist of $\omega_{r}=\SI{52.0}{\mu m}$ including the diamond. This is achieved by a symmetric laser cavity geometry with mirror radius of curvature $ROC=\SI{30}{mm}$ and geometric cavity length of $L=\SI{12}{mm}$. To address all NV-centres in the mode volume and still maintain a high power density the beam waist of the green laser $\omega_{g}=\SI{55.5}{\mu m}$, measured with a beam camera, is slightly bigger than the cavity mode. The high pump power enables pumping with a maximum power density of $\SI{0.1}{MW/cm^{2}}$ of the diamond medium mode volume of $\SI{0.003}{mm^{3}}$ corresponding to $\approx 10^{12}$ NV-centres.\par
Fig. 1b shows the energy level scheme of the NV-centre. The maximum gain from NV-centres is expected at the emission of the 3-phonon sideband at $\SI{710}{nm}$ \cite{Jeske2017, Fraczek2017,Albrecht2013, Su2010}. Additional seeding of the cavity is therefore realised by a laser at $\lambda_{seed}=\SI{710}{nm}$. To enable pumping and at the same time achieve high red intensity to observe stimulated emission, i.e. seeding the cavity, both lasers are combined with a dichroic mirror. Thus, the addressed third phonon-sideband transition (see indicated stim. em. transitions in Fig. 1b) corresponds to a 4-state laser level system at the peak NV spectrum emission \cite{Su2010}.\par
High gain in a laser cavity requires on the one hand a high number of emitters. On the other hand, the cavity losses at the lasing wavelength of $\lambda=\SI{710}{nm}$ need to be minimised, which leads to a high finesse and consequently many round trips of the emitted photons in the cavity. Thus, the diamond plate is polished on both sides to a surface roughness of $R_{a}=\SI{0.5}{nm}$ and placed in Brewster's angle to minimise the scattering and reflection losses at the diamond surface, respectively. Furthermore, we achieve an ultra-low absorption of the diamond host crystal of $\mu_{abs}=\SI{0.01}{cm^{-1}}$ at $\lambda=\SI{710}{nm}$ (see Supplementary). To further minimise the transmission losses of the cavity mirrors we choose two identical, highly reflective mirrors $R=\SI{99.98}{\%}$. This results in a poorly impedance matched cavity, however, we gain an improved ratio of the stimulated emission signal to the signal of the seeding laser transmitted through the cavity.\par
The finesse of the cavity is given by $\mathcal{F}=FSR/FWHM=\Delta\lambda/\delta\lambda$, where FWHM is the full width at half maximum and FSR the free spectral range of the resonances, see Fig. 2a. The FSR can be expressed as $FSR=c/(2L_{opt})$, which is constant when the seeding wavelength is fixed. Here, $c$ is the speed of light in vacuum and $L_{opt}$ the optical cavity length. If we assume a high finesse of $\mathcal{F}=800$, it follows $FWHM=\SI{15}{MHz}$. We use the seeding laser at $\lambda_{seed}=\SI{710}{nm}$ with a much smaller linewidth ($\SI{<100}{kHz}$), such that we can measure the FWHM by scanning the thin laser line over the resonance. To precisely measure the finesse of the cavity the laser is combined with a piezo that linearly scans the second mirror (M2) of the cavity over a maximum range of $\SI{3}{\mu m}$ (see Methods).
\begin{figure}
\centering
\includegraphics{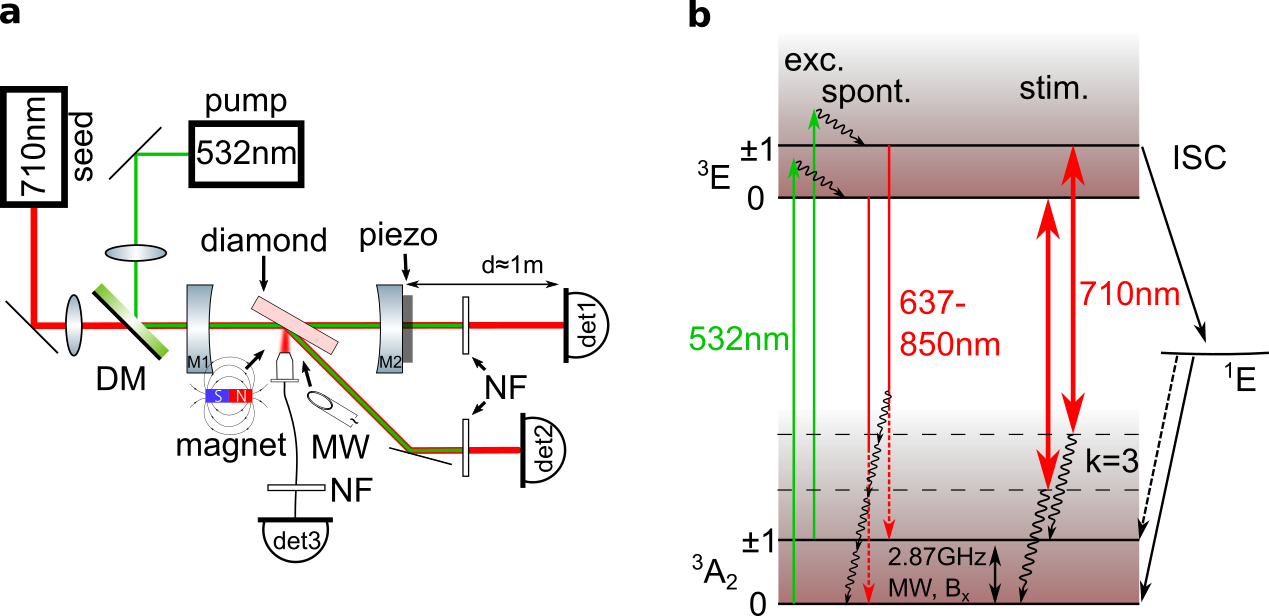}
\caption{\textbf{a}, Experimental setup: The pump laser ($\SI{532}{nm}$) and the seeding laser ($\SI{710}{nm}$) are combined with a dichroic mirror (DM) and individually focused into the cavity. The green laser is blocked via $\SI{532}{nm}$ notch filters (NF). Detected is the transmitted (det1), reflected (det2) and PL signal (det3). \textbf{b}, NV$^{-}$-centre energy level schematic. The system is pumped by a $\SI{532}{nm}$ laser from the ground ($^{3}A_{2}$) to the excited state ($^{3}E$). The spontaneous emission (spont. em.) spectrum is in the red to NIR regime ($\SIrange[range-phrase=-]{637}{850}{nm}$). The effective magnetic field $B_{x}$, as well as a continuous microwave drive at $\SI{2.87}{GHz}$ lead to spin mixing. Phonon levels are indicated by the shaded area. The 3-phonon transition of the spin $m_{s}=0$ state at $\lambda=\SI{710}{nm}$, where stimulated emission (stim. em.) occurs, is indicated with the large double arrow. The transition for the degenerated spin $m_{s}=\pm1$ states is left out for visibility reasons.}
\label{fig1:setup_NV_level_scheme}
\end{figure}

\subsection{Stimulated emission sensing}
\label{subsec:Stimulated emission sensing}
The NV-diamond-loaded, high-finesse cavity setup allows us to study the amplification behaviour when the green laser pumps the NV-centres. Fig. 2a shows that we achieve a high-finesse cavity of $\mathcal{F}=710$. In the case of a high finesse the Airy function can be approximated by a Lorentzian \cite{Pollnau2020}. The finesse of a laser cavity below lasing threshold that contains a gain medium is given by $\mathcal{F}=\pi/(\mu_{0}L - \mu_{g}l)$ \cite{Siegman1986}, where $\mu_{0}$ is the loss per unit length of the cavity when only the red seeding laser is used, $L$ is the geometric cavity length, $l$ is the geometric length of the gain medium and $\mu_{g}$ is the gain per unit length from the NV-centres in the diamond. A change in gain can therefore be observed in the FWHM and finesse of the cavity if the seeding wavelength is constant. It is also distinct from thermal influences on the cavity that would affect the finesse via a change in the FSR. When pumping the NV-centres we see that the additional gain by stimulated emission reduces the loss, consequently an increase in the finesse is expected \cite{Siegman1986}.\par
The green trace in Fig. 2a shows the detection of the amplification by stimulated emission of the cavity modes. The seeding power was set to $P_{seed}=\SI{7.7}{mW}$. By pumping the NV-centres we detect a clear amplitude amplification of $\SI{45}{\%}$ of the resonances which corresponds to a raise in the cavity output power from $\SI{3.0}{\mu W}$ to $\SI{5.5}{\mu W}$. In addition, the increase of the finesse is caused by the reduction of the FWHM, as shown in the inset of Fig. 2a. When only the green laser is pumping the NV-centres we detect a flat signal. This verifies that the detected amplification is not caused by the spontaneous emission of the NV centres.\par
Fig. 2b shows the behaviour of the amplification as a function of the green pump power by detecting the amplitude of the cavity mode A$_{\text{cavity}}$. The maximum of the amplification is at $\approx\SI{1}{W}$, corresponding to an intensity of $\SI{103}{W/mm^{2}}$. We investigate the origin of the amplification by the detection of the cavity FSR and FWHM during a scan of the pump power (see Supplementary). We find that the amplification comes from a reduction of the FWHM, that is, a change in the net gain of the NV-centres. The gain also leads to a higher number of photons in the cavity, as expected from stimulated emission. The FSR is constant when the pump power is increased. This proves that no thermal expansion is influencing the finesse and amplitude, as $FSR\propto L_{opt}$. \par
To exclude that the increase of the finesse comes from a reduction of the initial absorption of the sample, we investigate samples with an absorption between $\SIrange[range-phrase=-]{0.3}{0.01}{cm^{-1}}$ and find out that the strongest amplification comes from the sample with lowest absorption (see Supplementary), which we use for all other measurements. Calculations of the NV ground-state 3-phonon level population in combination with the absorption and amplification indicate that the initial absorption of the NV-centre with a wavelength of $\SI{710}{nm}$ does not play a significant role (see Supplementary). As the initial absorption of the used sample in our experiments has an ultra low absorption of $\SI{0.01}{cm^{-1}}$ this shows that we see stimulated emission and not a reduction in initial absorption.\par
Gain in the cavity is directly detected by the amplification of the amplitude $\text{A}$, which is proportional to the detected power and intra cavity power. The amplitude which is the maximum of the cavity resonance would correspond to the power output of a cavity locked to its maximum. We use the cavity amplitude of the transmitted $\text{A}_{\text{trans}}$ (det1) or reflected $\text{A}_{\text{ref}}$ (det2) signal for further quantification of the cavity behaviour. \par
The net gain from the NV-centres is shown Fig. 2b (red trace). It is calculated from the finesse by $\mu_{g}=\pi/l(\mathcal{F}_{0}^{-1}-\mathcal{F}_{g}^{-1})$ (see Supplementary). Here, $\mathcal{F}_{0}$ is the finesse at zero pump power. The maximal net gain is $\SI{2.04}{m^{-1}}$ at $\approx\SI{1}{W}$. For higher pump powers the gain, as well as the amplitude decrease. At a pump power above $\SI{3}{W}$ both curves decrease below their initial value at zero pump power, ending the regime of net gain. The high pumping power regime above $\SI{3}{W}$ shows a linear increase of the losses (FWHM) (see Supplementary). This reveals a new effect of induced absorption at the wavelength of $\SI{710}{nm}$ in the diamond caused by the green pumping laser, which was likely a major hurdle in previous attempts at lasing \cite{Nair2020, Fraczek2017, Webb2021}. The data shows that this new effect of induced absorption compensates the gain by stimulated emission at a pumping power of $P_{532}\approx\SI{3}{W}$.\par
The amplification is dependent on the pump and seeding power as shown in Fig. 2c. Here the amplification $\Delta\text{A}=(A-A_{0})/A_{0}$ is calculated from the relative difference between the amplified cavity amplitude $\text{A}$ and the amplitude without pumping $\text{A}_{0}$. The measurement shows an optimal power regime for the amplification by stimulated emission. The highest amplification by stimulated emission is at $P_{532}=\SIrange[range-phrase=\,\text{to}\,]{1}{2}{W}$ and small red seeding powers $P_{710,in}<\SI{0.2}{W}$. In this regime the amplification by stimulated emission reaches $>\SI{64}{\%}$ which is the strongest light amplification by NV-centres detected so far.

\begin{figure}
\centering
\includegraphics{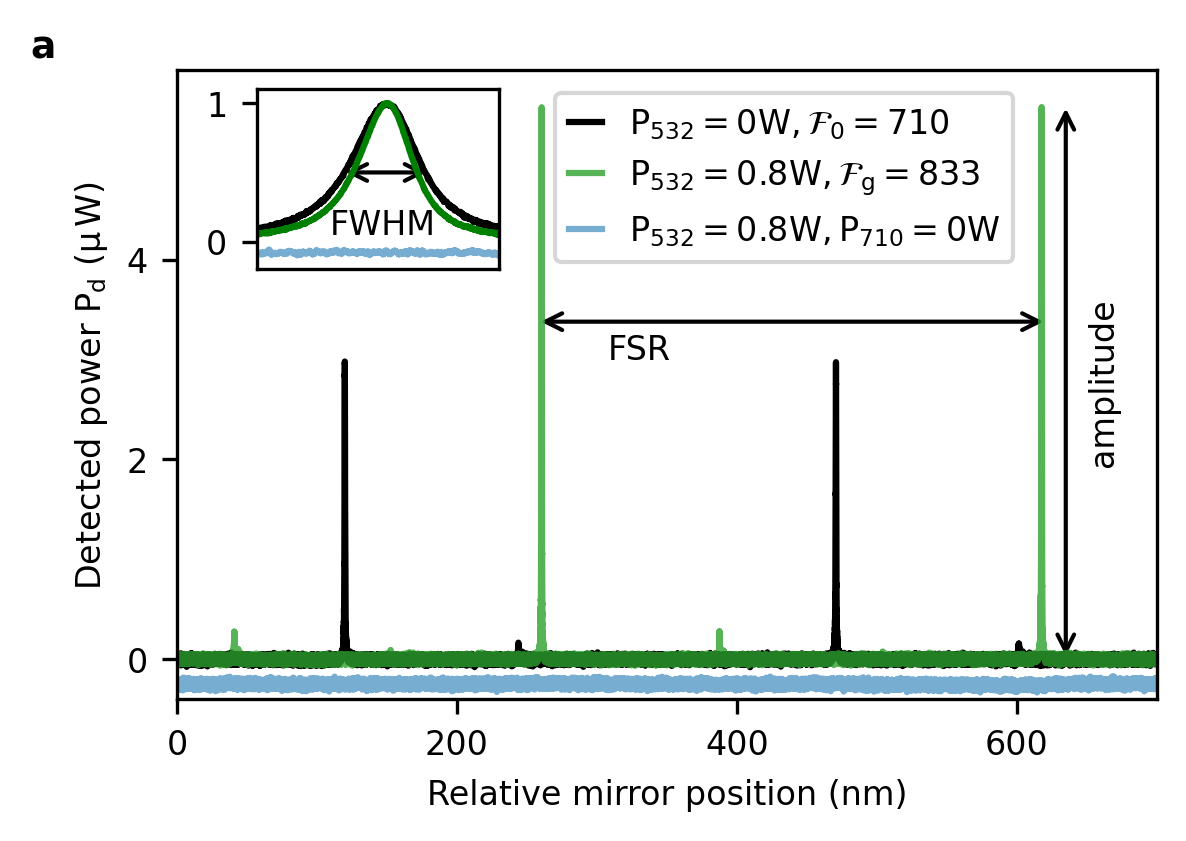}
\includegraphics{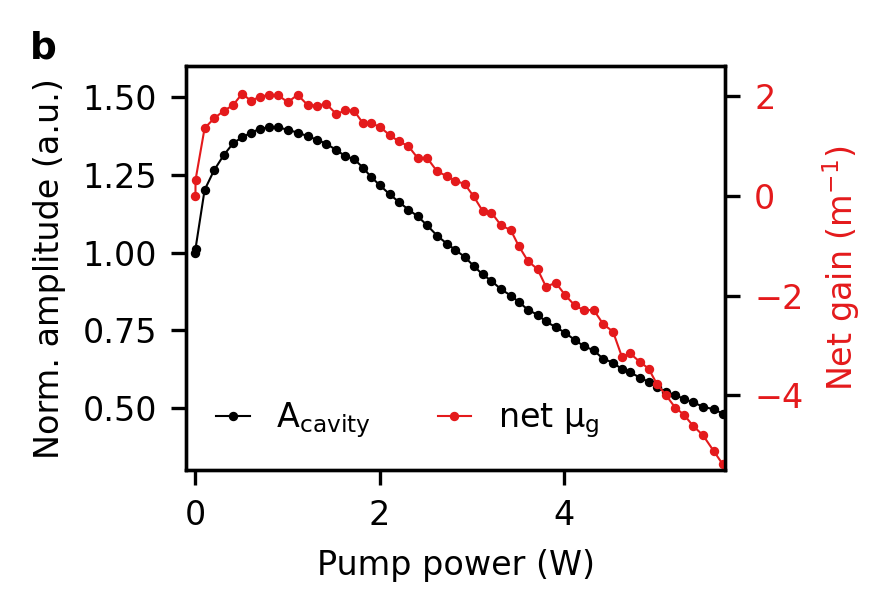}
\includegraphics{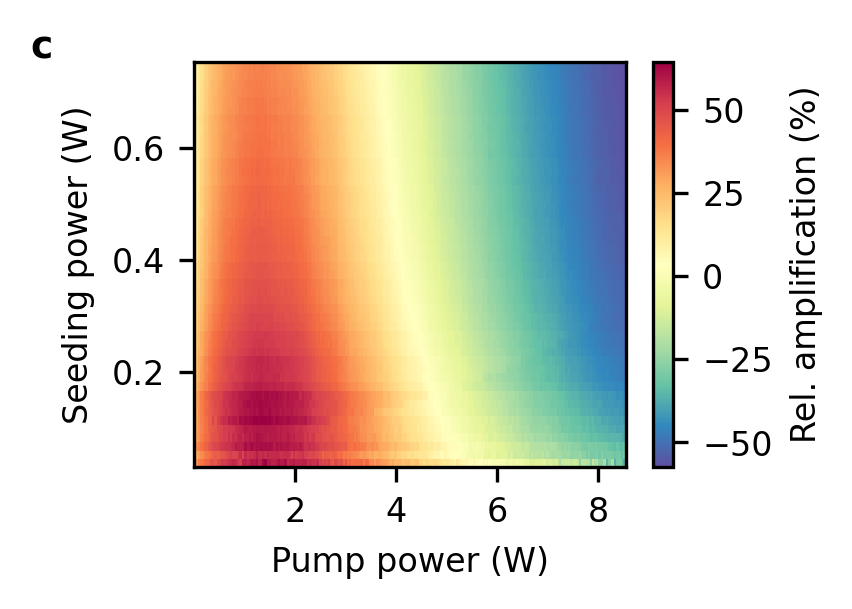}
\caption{\textbf{a}, Finesse and amplification measurement of the cavity including the diamond. The black trace shows the signal of the seeding laser only at $\SI{710}{nm}$. When the green laser pumps the NV-centres, the finesse and amplitude increase (green trace). The signal of only the green laser pumping (blue trace) is shifted for better visibility. The inset shows a zoom into normalised and overlapped peaks. \textbf{b}, Cavity amplitude A$_{\text{cavity}}$ (black) and net gain $\mu_{g}$ (red) as a function of the pump power. A$_{\text{cavity}}$ is normalised to the amplitude at zero pump power. The net gain is calculated from the measured finesse (see Supplementary). \textbf{c}, Amplification $\Delta A$ over input pumping power $P_{532}$ and input seeding power $P_{710,in}$. The amplification $\Delta A$ is the relative difference to the case of zero pumping power $P_{532}=\SI {0}{W}$. Maximal amplification of $\SI{64}{\%}$ occurs at $\SI{1}{W}$ to $\SI{2}{W}$ pump and $\SI{<0.2}{W}$ seeding power.}
\end{figure}

%############################################################################################################################################################
\subsection{Magnetic field dependency}
\label{subsec:Magnetic field dependency}
After the demonstration of strong light amplification we investigate the magnetic field dependency of the measured amplification. As only the emission of the NV-centre decreases with an external magnetic field, this proves that the amplification is caused by the NV-centres.  We particularly focus on the question whether the competition between stimulated and spontaneous emission, i.e. the non-linearity of the cavity leads to a signal amplification, which should show up via increased contrast compared to spontaneous emission.\par
We performed the measurements on the same setup as described in Fig. 1a. To apply a homogeneous magnetic field inside the diamond we use as a first experiment a strong permanent magnet. The magnet is guided from the top by a linear piezo stage (shown in Fig. 1a from the side for purposes of representation). The magnetic field is applied in the (100) crystal direction. The diamond contains NV-centres in all four possible NV-directions. In this configuration the projections of the applied magnetic field onto the perpendicular plane to the NV-axis is symmetric and of the same magnitude for all NV-orientations. The angle between the NV-axis is $\alpha=\SI{109.5}{\degree}$ \cite{Doherty2013}. The effective magnetic field that the electron spin is interacting with is then calculated by $B_{x}=B\cos(\alpha/2)$, where $B$ is the magnitude of the measured magnetic field. We assume a constant magnetic field within the mode of the diamond of diameter $2\omega_{0}\approx\SI{110}{\mu m}$. The excited NV-centres in the diamond mode in the experiment are located at a distance of $\SI{2\pm1}{mm}$ from the magnet. The effective magnetic field is $B_{x}=\SI{105\pm25}{mT}$ (details are shown in the Supplementary).\par
Fig. 3a shows measurements of the cavity TEM$_{00}$ modes in different conditions. The cavity mode excited by the red laser gives a finesse of $\mathcal{F}=\SI{958\pm6}{}$ and is not influenced by the magnetic field. The coherent stimulated emission when pumping the NV-centres couples into the cavity mode and amplifies the amplitude of the resonances. The result leads to an amplification of $\Delta A=\SI{30}{\%}$ and a finesse of $\mathcal{F}=\SI{1086\pm2}{}$. The presence of a magnetic field suppresses the amplification significantly and leads to a finesse of $\mathcal{F}=\SI{926\pm3}{}$. The decreasing finesse shows that there is a reduction in gain in the cavity when a magnetic field is applied. The reduction is due to the magnetic field shifting population from the "brighter" $m_{s}=0$ state to the "darker" $m_{s}=\pm1$ states. These "darker" $m_{s}=\pm1$ states show an intersystem crossing (ISC) to the long living singlet state ($^{1}$A), depicted in Fig. 1b. The measurements demonstrate the first magnetic-field dependent amplification by stimulated emission from NV-centres.\par
In fact the reduction of the signal is equal to the entire amplification, suggesting that due to the cavity enhancement the entire signal from the NV centres is reduced by the magnetic field. This opens the potential to achieve close to $\SI{100}{\%}$ contrast with lock-in detection where the NVs emission is modulated via MW and allows to filter out only the NV emission and reject the unmodulated seeding laser signal. In this work we analyse contrast only on the basis of the combined signal from NV centres and seeding laser.\par
Furthermore, we investigate the contrast caused by the magnetic field dependent stimulated emission. The amplitude of the cavity mode is detected while scanning the pump and seeding power, once with and once without the magnet close to the NV-centres. The optimal power regime for the relative detected contrast is shown in Fig. 3b. The contrast is calculated from $Contrast=(A-A^{mag})/A$, where $A^{mag},A$ are the detected amplitude with and without a magnet, respectively. The contrast reaches a maximum value of $\SI{32.6\pm0.4}{\%}$. The range above $\SI{25}{\%}$ is at pump powers between $\SIrange[range-phrase=-]{2}{6}{W}$ over a large red seeding power regime of $\SIrange[range-phrase=-]{10}{650}{mW}$.\par
If we compare the optimal amplification regime in Fig. 2c to the optimal pump power for maximal contrast, the optimal contrast is at higher pump powers than the optimum for amplification. We explain the difference by the interplay between the amplification by stimulated emission and the induced absorption discussed in the previous section. We have found that the induced absorption is dependent on the pump power, that is, on the excitation rate of the NV-centres. The gain $\mu_{g}$ by stimulated emission depends on the excitation rate, as well, as it is proportional to the excited state population $\mu_{g}\propto\rho_{exc}$ \cite{Milonni2010}. To optimise the magnetic field dependency the NV-centres have to be spin-polarised to the $m_{s}=0$ ground state. The spin polarisation is also dependent on the excitation rate. As the beam waist in the focal spot is $w_{g}=\SI{55.5}{\mu m}$, high pump powers are needed for sufficient spin polarisation. We see from the measurements that at the point of maximal amplification, which is an interplay between the stimulated emission and induced absorption, the spin polarisation is not high enough to reach the highest magnetic field response and contrast. That point is reached at higher excitation rates where the induced absorption is already compensating the stimulated emission.\par
Final verification of the enhanced magnetic-field-dependent response of the NV-centres is done by a direct comparison between the magnetic-field-dependent PL  and coherent cavity output (det3 and det1 in Fig. 1a, respectively) while a transverse magnetic field is introduced and removed, which mixes the spin states and reduces the NV signal. We use a pump power $P_{532}=\SI{3.4}{W}$ and seeding power $P_{532}=\SI{25}{mW}$ in the regime of high contrast, of Fig. 3b. In Fig. 3c the NV-centres PL is detected over time. The measured contrast of the PL reaches $\SI{15.91}{\%}$ which is a high contrast when measuring the PL of NV ensembles. It is reduced due to inhomogeneous broadening and background PL (e.g from NV$^{0}$) \cite{Rondin2014}. On the basis of the standard NV energy level scheme presented in \cite{Doherty2011} the maximal contrast that can be reached with a permanent magnet is $\SI{22}{\%}$ (see Supplementary).\par
Simultaneously, the amplitude of the cavity resonance is detected in Fig. 3c. The contrast of the cavity mode ($A_{\text{cavity}}$) is $\SI{32.65}{\%}$, that is, more than twice as high as the contrast detected by collecting only the PL of the NV-centres. Our coherent cavity readout surpassed the fundamental contrast which is possible via PL collection and, to our knowledge, any contrast measured with single NV-centres. The measurement shows the higher contrast because of the competition between spontaneous and stimulated emission in the cavity. The magnetic-field-induced reduction of gain results in a decreased NV-centre emission which is enhanced by the non-linear cavity effect.\par
Fig. 3d shows the magnetic-field-dependent behaviour of the transmitted and reflected signal amplitude over the pump power. The magnetic field influences the transmitted and reflected signal in the same way. The maximum of the amplitude is shifted to lower pump powers in the presence of a magnetic field compared to no magnetic field and a reduction of the amplitude with magnetic field over the complete pump power range is detected. The magnetic field dependency of the amplitude remains strong also at high pump powers above $\SI{2}{W}$ where there is no net gain. This confirms again induced absorption by the green pump laser and the interplay between stimulated emission and the induced absorption as discussed previously in this section. The magnetic field reduces the gain in addition to the induced absorption.\par
As a next step we investigate the potential for high signal output of the system. The magnetic field sensitivity of the system is improved by both an increased contrast and a high detected signal power \cite{Degen2017}. Due to the highly reflective mirrors the transmitted signal of the cavity in our configuration is weaker than the reflected signal at the diamond surface (see det3 in Fig. 1a). The reflected signal is caused by birefringence losses turning the p-polarised light without reflection losses under Brewster's angle to partially reflected s-polarised light. Because of the low absorption sample these losses become significant.\par
The detection of the reflected and transmitted signal is shown in Fig. 3d. The data shows that the detected signal for the transmitted amplitude is at $\approx\SI{1}{mW}$. By detecting the reflection the signal enhances to $\approx\SI{8}{mW}$. This enables the detection of stimulated emission signals from an NV ensemble with a mW strength. Thus, by using the cavity laser output we increase the signal strength significantly compared to the PL detection which has  been used for sensing applications up to date. We point out that the cavity signal exits as a beam, and can be directed to a distant detector (see. Fig. 1a), while the PL requires collection via a microscope objective close to the sample.

\begin{figure}
\centering
\includegraphics{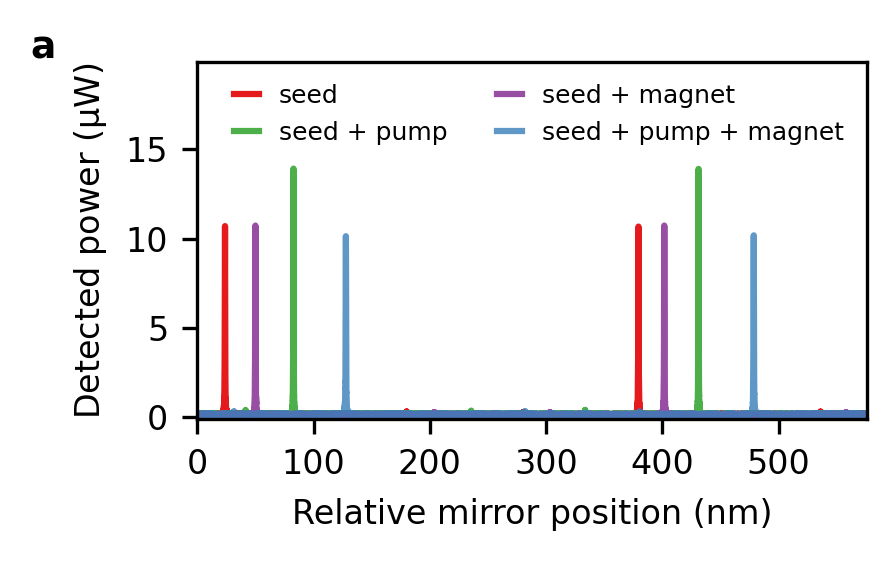}
\includegraphics{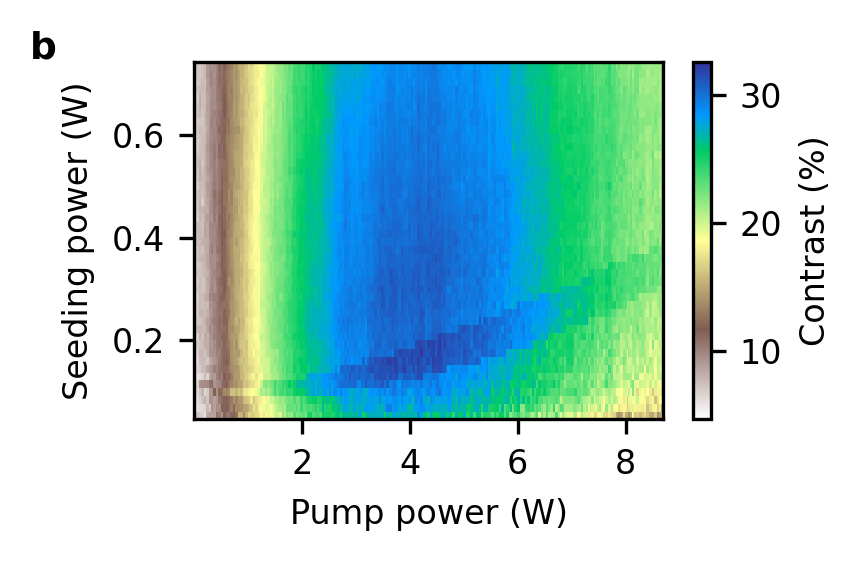}
\includegraphics{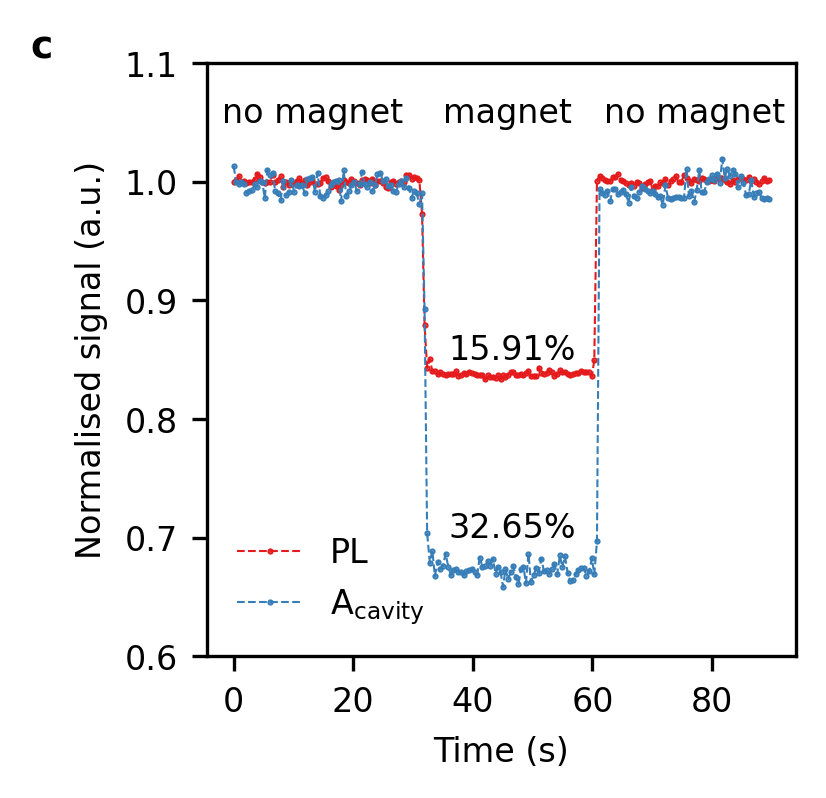}
\includegraphics{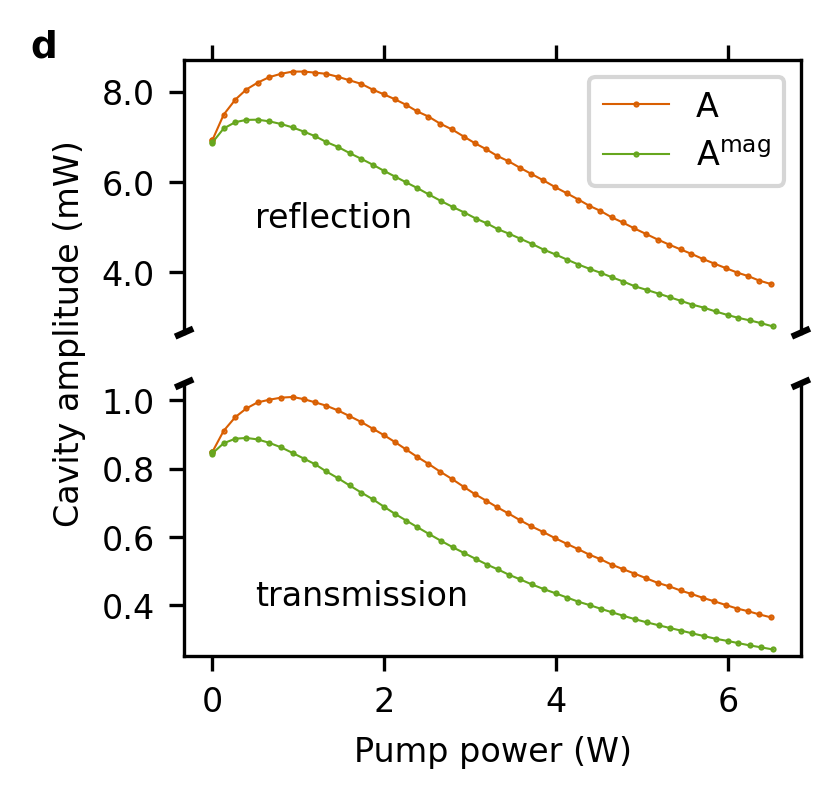}
%\hfill
%\subfigure[]{\label{fig:b_amp_modes_sim}\includegraphics[width=70mm]{simulation_modes_r_rg_rgmag_g}}
\caption{\textbf{a}, Detection of the transmitted cavity power $P_{\text{d}}$ over the relative mirror position. The transmission of only the red laser (seed) is not influenced when a strong magnet is brought close to the diamond (seed + magnet). Strong amplification of the mode appears when the green laser pumps the NV centres (seed + pump). Bringing a magnet close to the diamond significantly reduces the amplified signal (seed + pump + magnet). The power for the seed and pump laser is $P_{710}=\SI{0.3}{W}$ and $P_{532}=\SI{1.37}{W}$, respectively. \textbf{b}, Contrast of the detected amplitude over the pump $P_{532}$ and seeding $P_{710,in}$ power. \textbf{c}, Transmitted peak amplitude (det1) of the cavity mode $A_{\text{cavity}}$ and laterally emitted PL (det3) are detected simultaneously over time. When the magnet is brought close to the diamond for $\SI{30}{s}$ both signals decrease. \textbf{d}, Measurement of the reflected (det2) and transmitted (det1) amplitudes in absolute values with (A$^{\text{mag}}$) and without (A) magnet over the pump power $P_{532}$ (note the subdivided vertical axis). The seeding power is set to $P_{710}=\SI{1.14}{W}$. The data represent the mean value of eight measurements.}
\end{figure}

%############################################################################################################################################################
\subsection{ODMR and sensitivity measurement}
So far we have demonstrated an increased contrast of the NV-centres by a constant permanent magnetic field. To predict sensitivities of the current setup, we demonstrate optically detected magnetic resonance (ODMR) measurements in a second experiment. To apply the microwaves, a loop antenna is brought close to the pumped NV-centres in the cavity mode from one side of the diamond as depicted in Fig. 1a. We expect to see an increased contrast and higher detection signal output in the cavity readout, compared to the PL measurement. The magnetic field sensitivity should be improved compared to conventional NV-centre ODMR.\par
Fig. 4a shows the ODMR measurement of the transmitted (det1) and PL signal (det3). The pump and seeding powers were $P_{532}=\SI{2.78}{W}$ and $P_{710}=\SI{1.51}{mW}$, respectively. The overall contrast of the ODMR PL signal is $\SI{11.4}{\%}$ and for our cavity setup $\SI{17.4}{\%}$. We fitted a double Lorentzian to the data. The parameters for the single Lorentzian with maximal contrast were $C_{\text{PL}} = \SI{8.9\pm0.0}{\%}$ and $C_{\text{A}_{\text{cavity}}} = \SI{13.6\pm0.1}{\%}$ with a corresponding ODMR line width $\Delta\nu_{\text{PL}} = \SI{5.63\pm0.02}{MHz}$ and $\Delta\nu_{\text{A}_{\text{cavity}}} = \SI{4.46\pm0.04}{MHz}$, for the PL measurement and transmitted cavity signal, respectively. As expected the contrast of our cavity setup exceeds the PL contrast.  In addition, the ODMR line width is decreased compared to the PL measurement, as well. We explain the reduced contrast in both cases compared to the permanent magnetic field contrast by an inhomogeneous distribution of the magnetic field amplitude over the diamond medium. The achieved record contrast in \ref{subsec:Magnetic field dependency} is an indication of the achievable ODMR contrast, when the microwave delivery is achieved homogeneously over the cavity volume.\par
The ODMR measurement allows us to determine and directly compare the sensitivity that is reached by a PL measurement and our cavity enhanced sensing. The shot-noise limited DC sensitivity based on optically detected magnetic resonance (ODMR) measurements from NV-centres is $\eta\propto \Delta\nu/(C\sqrt{I_{0}})$ \cite{Rondin2014, Dreau2011}. The sensitivity can therefore be improved by narrowing the linewidth $\Delta \nu$ of the ODMR or increasing the contrast $C$ or detected signal $I_{0}$. The output of the cavity readout can be increased with the seeding power and Fig. 3b indicates that the contrast only reduces slowly, while the signal can be strongly enhanced. For improved sensitivity, we therefore go into a regime of higher seeding power above $P_{710}=\SI{1}{W}$. The measurements are shown in Fig. 4b. We determine the sensitivities from the fit parameters of the ODMR (see Supplementary). The resulting shot-noise limited DC magnetic field sensitivity of the PL measurement is $\eta_{\text{PL}} = \SI{135.5\pm2.4}{pT/\sqrt{Hz}}$. For our coherent cavity readout the dc sensitivity is $\eta_{\text{cavity}} = \SI{14.6\pm1.3}{pT/\sqrt{Hz}}$, that is, an improvement of one order of magnitude.

\begin{figure}
\centering
\includegraphics{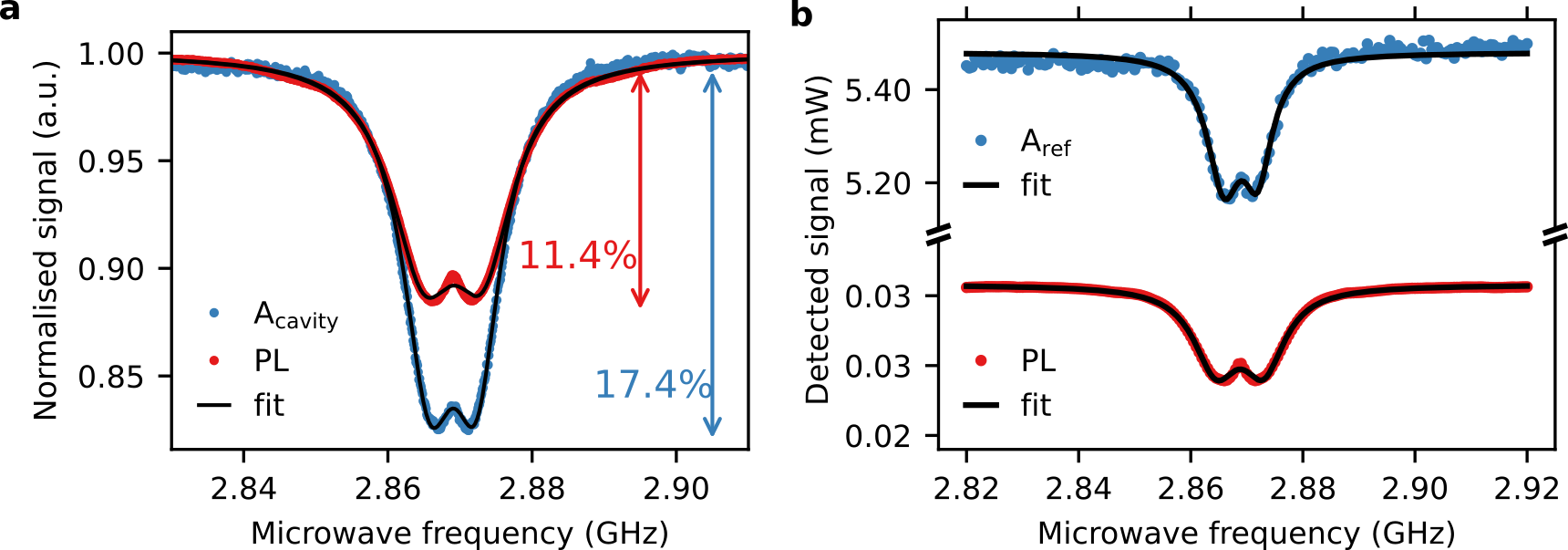}
\caption{ODMR measurements optimised for contrast \textbf{a} and sensitivity \textbf{b}: \textbf{a}, Normalised detected power of the transmitted cavity amplitude (A$_{\text{cavity}}$, det1) and PL (det3) signal as a function of the microwave frequency. The arrows show the overall contrast of the measurement. \textbf{b}, Detected absolute power of the reflected cavity amplitude (A$_{\text{ref}}$, det2) and PL signal. A double Lorentzian is fitted to the data for both cases (black line).}
\end{figure}

%############################################################################################################################################################
\section{Discussion}

We have shown light amplification by stimulated emission from NV-centres of $>\SI{64}{\%}$ in a macroscopic laser cavity with an optimal pump power of $\SIrange[range-phrase=-]{1}{2}{W}$ at a red seeding power of $<\SI{0.2}{W}$. We have also achieved stimulated emission signals from NV centre ensembles with an abolute output power in the mW range. This is both the highest relative amplification and the highest absolute signal of stimulated emission from NV centres measured so far. We show that induced absorption by the green pump laser is an additional loss at the red lasing wavelength of $\SI{710}{nm}$ that leads to a decrease of the amplification at high pump powers $>\SI{2}{W}$. This additional loss path is identified as a purely material-related absorption loss induced by the green laser since the linewidth and finesse are reduced, i.e. loss increased while the free spectral range is constant. Possible candidates could be ionisation to the conduction band and tunneling of the electron to other defects (e.g. nitrogen or nitrogen complexes) leading to a reduction in quantum efficiency and the creation of new charge states of defects which absorb at this wavelength.\par
Furthermore, we detect magnetic field dependency of the amplification by stimulated emission of the NV-centres. The resulting contrast in the amplitude of the cavity resonance is detected to be $\SI{26}{\%}-\SI{33}{\%}$ at an output power between $\SIrange[range-phrase=-]{0.1}{6}{mW}$, depending on the detection of the transmitted and reflected signal of the cavity. The achieved contrast is a new record for an ensemble of NV-centres and is higher than what can possibly be achieved with spontaneous emission.
We demonstrate experimentally the advantages of coherent cavity readout for sensing and the principle of laser threshold magnetometry for the first time. This opens the door for a variety of novel sensing techniques and applications, precision improvements in sensors and the exploration of coherent readout of quantum systems for sensing, quantum bits and quantum technology more generally.\par
The detection of ODMR with coherent laser output shows a magnetic field sensitivity of $\SI{14.6\pm1.3}{pT/\sqrt{Hz}}$ which is an improvement of one order of magnitude compared to the conventional PL readout.\par
For further improvement and the development of a highly sensitive magnetic field sensor, a sensitivity measurement can be done via locking the cavity. The ODMR measurement can then be combined with a CW signal detection. An increase in the signal output is expected by impedance matching the mirrors. In addition via lock-in detection and MW modulation of the NV emission the red seeding laser could be filtered out and amplification and contrast measurements indicate that then a contrast of almost unity might be reached. In addition, the development of a microwave antenna, e.g. a loop antenna on both sides of the diamond or a microwave resonator, could lead to a more homogeneous microwave field and an increased ODMR contrast.

%############################################################################################################################################################
\section{Methods}
\subsection{Diamond material}
The sample used for our studies is a (100) oriented, both sides polished, commercial HPHT diamond from ElementSix with an edge length of $\SI{3}{mm}$ and a thickness of $l=\SI{295}{\mu m}$. It is treated with LPHT annealing at $\SI{1800}{\degree C}$ in an inert gas atmosphere to reduce the initial absorption before electron irradiation with a fluence of $\SI{1e18}{cm^{-1}}$ and a electron energy of $\SI{2}{MeV}$. After irradiation the sample was again annealed at $\SI{1000}{\degree C}$ for two hours to create NV-centres.
\subsection{Experimental setup}
For all measurements we used the setup in Fig. 1a. We used lenses to prepare the beam size for optimal mode matching of the seeding laser and matching the pump beam to be slightly larger than the cavity mode. To efficiently couple into the cavity mode and pump the NV centres we horizontally polarised both beams as the diamond was placed in Brewster's angle. The cavity length was modulated by applying a sawtooth signal to the piezo with a frequency of $\SI{111}{Hz}$ and an amplitude of $\SI{3}{\mu m}$. Several fundamental modes of the cavity were sampled with an oscilloscope. The piezo was scanned in closed loop mode, to detect the position. The cavity resonances were detected in the linear regime of the piezo movement. The power of the transmitted (det1) and reflected (det2) cavity signal was detected through variable gain photo receivers. The photoreceivers were power calibrated. Immediate analysis by integrated oscilloscope functions enabled a fast read out of the measurement quantities, i.e. the cavity resonance amplitude, FWHM and FSR.\par
The power of the PL (det3) was directly measured by a photo-diode power sensor. The use of the microwave antenna for the ODMR measurement lead to a decrease in the PL signal because of partially blocking the PL light. To have comparable PL measurements we used a saturation measurement of the PL without microwave antenna where the PL collection was maximised. We used this measurement as a calibration of the PL signal (see Supplementary).

\newpage

%###########################################################################################################
%###########################################################################################################
%% --------------------
%% |   Bibliography   |
%% --------------------
\cleardoublepage
\phantomsection
\addcontentsline{toc}{chapter}{\bibname}

\bibliographystyle{unsrt}											  
\bibliography{library}

%############################################################################################################################################################
\section{Acknowledgements}
F.H. and J.J. acknowledge funding from the German federal ministry for education and research, Bundesministerium für Bildung und Forschung (BMBF) under Grant No. 13XP5063.
Part of this study was carried out within the framework of the QST International Research Initiative (QST IRI).
B.C.G. and A.D.G. acknowledge funding from the United States Office of Naval Research Global (ONRG) Global-X Challenge (N62909-20-1-2077), Asian Office of Aerospace Research and Development (FA2386‐18‐1‐4056) and the Australian Research Council (ARC) Centre of Excellence for Nanoscale BioPhotonics (CE140100003). A.D.G. acknowledges financial support from the Australian Research Council under the Future Fellowship scheme (FT160100357). We thank Philipp Reineck and Marko Härtelt for valuable discussions and insights.

%############################################################################################################################################################
\section{Author Contributions}
F.H. and J.J. conceived the idea of the cavity setup, F.H. prepared the experimental setup, performed the experiments and analysed the data, L.L. and X.V. supported in building the experimental setup and with helpful discussions about the setup development. A.M.Z. performed the LPHT annealing of the sample. T.O., S. O. and S. I. performed electron irradiation and annealing for NV-creation of the sample. T.L. helped characterising the samples, F.H., J.J., L.L., X.V., A.D.G., B.C.G. and M.C. discussed and interpreted the results and provided feedback to the manuscript. F.H. and J.J. did the calculations, F.H. wrote the manuscript with the support of J.J..

\end{document}